\documentclass[twocolumn]{autart}    % Enable this line and disable the 
\newtheorem{problem}{{\bf Problem}}
\newenvironment{proof}
  {\par\noindent\textit{Proof.}\ }  % Inizio con "Proof." in corsivo
  {\hfill$\blacksquare$\par}            % Fine con quadratino e nuova riga
\newtheorem{proposition}{Proposition}
\newtheorem{assumption}{{\textit{Assumption}}}
\newtheorem{remark}{{\textit{Remark}}}

\usepackage{amsmath, amssymb, amsfonts, mathtools} % math symbols and sets
\usepackage{algorithm, algpseudocode} % environment for algorithm
\usepackage{graphicx, soul}
\usepackage{lipsum} 
\usepackage{color}
\usepackage{subcaption} % for side-by-side figure arrangement

\usepackage[authoryear]{natbib}
\usepackage[colorlinks=true,linkcolor=blue,citecolor=blue,urlcolor=blue]{hyperref}

\usepackage{graphicx}
\usepackage{epstopdf}
\begin{document}

\begin{frontmatter}

\title{ Distributed Unknown Input Observers for Discrete-Time Linear Time-Invariant Systems\thanksref{footnoteinfo}} % Title, preferably not more 

\thanks[footnoteinfo]{This paper was not presented at any IFAC 
meeting. Corresponding author F.~Tedesco.}

\author[DIMES]{Franco Angelo Torchiaro}\ead{franco.torchiaro@dimes.unical.it},    % Add the 
\author[DIMES]{Gianfranco Gagliardi}\ead{g.gagliardi@dimes.unical.it},               % e-mail address 
\author[DIMES]{Francesco Tedesco}\ead{ftedesco@dimes.unical.it}, 
\author[DIMES,DIIES]{Alessandro Casavola}\ead{a.casavola@dimes.unical.it}  % (ead) 
              
\address[DIMES]{Department of Computer, Modeling, Electronics and Systems Engineering (DIMES), University of Calabria, Rende (CS), Italy 87036}             % full addresses
\address[DIIES]{Department of Information Engineering, Infrastructure and Sustainable Energy (DIIES), Mediterranea Univeristy of Reggio Calabria, Reggio Calabria (RC), Italy 89124}        % here.

                             % keyword list or with the 
                                          % help of the Automatica 
                                          % keyword wizard

\begin{abstract}                          % Abstract of not more than 200 words.
This paper introduces a Distributed Unknown Input Observer (D-UIO) design methodology that uses a technique called node-wise detectability decomposition to estimate the state of a discrete-time linear time-invariant (LTI) system in a distributed way, even when there are noisy measurements and unknown inputs.
In the considered scenario, sensors are associated to nodes of an underlying communication graph. Each node has a limited scope as it can only access local measurements and share data with its neighbors. The problem of designing the observer gains is divided into two separate sub-problems: (i) design local output injection gains to mitigate the impact of measurement noise, and (ii) design diffusive gains to compensate for the lack of information through a consensus protocol.
A direct and computationally efficient synthesis strategy is formulated by linear matrix inequalities (LMIs) and solved via semidefinite programming. Finally, two simulative scenarios are presented to illustrate the effectiveness of the distributed observer when two different node-wise decompositions are adopted.
\end{abstract}

\end{frontmatter}

\section{Introduction}
\label{sec:introduction}
The growing prevalence of large-scale complex systems, including power grids and transportation networks, alongside multi-agent and networked systems, has rendered distributed state estimation a prominent and challenging area of research \citep{RPAJ:19}.
In a general distributed estimation setup \citep{MS:16}, a central plant is monitored by a network of sensor nodes, each equipped with computational and data exchange capabilities. The objective of each node is to asymptotically estimate the state of the dynamical system despite: (i) having only limited knowledge of the state, and (ii) communication being restricted to immediate neighboring nodes. 
In this context, centralized solutions are not feasible \citep{O:05}, making cooperation among nodes essential for successfully reconstructing the system state.

Considering that each agent's measurement of the plant’s state may suffer from a lack of detectability in relation to the plant dynamics, the authors in \citep{KSC:16} propose to address this issue through the intuition of node-wise detectability decomposition. This key idea contributed to a unique design of the distributed observer, where the local output injection gain and the neighboring coupling gain were distinctly associated with two separate components of the distributed observer.
In the subsequent work \citep{KLS:19}, the concept of decoupling the detectable part from the undetectable part led to a fully decentralized observer design strategy, which resulted into a plug-and-play distributed observer with adaptive coupling gains.
Building on the methodology of node-wise decomposition, authors in \citep{HTWS:18} developed an algorithmic procedure to design a suitable distributed observer, leveraging linear matrix inequalities (LMIs) to ensure computational efficiency.
More recently, in \citep{CW:23}, the same idea was applied to the design of a distributed Unknown Input Observer (UIO), where a closed-form solution was derived for the computation of the coupling gains. The problem of distributed state estimation in the presence of unknown inputs at some nodes was also addressed in \citep{YBRP:22} using a network of distributed observers. An additional interesting aspect discussed in \citep{YBRP:22} is the use of the UIO to overcome a common practical limitation: when a system is distributed and driven by external inputs, it may not be possible for each node to access all control signals. In this context, the unavailable local input data can be regarded as an additional unknown input.

A key advantage of the UIO is its ability to decouple error evolution from the unknown input without requiring specific assumptions on the behavior of the input, ensuring the observer's resilience to abrupt changes and extreme variations in the exogenous signal \citep{CW_R:23}.
All the previously mentioned solutions apply to the continuous-time domain, while limited results exist in the literature that exploit the node-wise detectability intuition for the discrete-time domain. Some of the challenges in adapting continuous-time solutions to discrete-time state estimation were explored in \citep{WLMA:19}, where a distributed observer was successfully developed for communication networks with time-varying topology. Specifically, a major challenge arises from the fact that existing design techniques proposed in continuous-time rely on the concept of ``high gain", which has no direct equivalent in discrete-time, presenting a significant obstacle.

Moving from this considerations, this work leverages node-wise detectability decomposition to elaborate a simple strategy for designing a Distributed Unknown Input Observer for discrete LTI systems. The proposed approach, based on linear matrix inequalities (LMIs) conditions, effectively handles unknown inputs and measurement errors while eliminating the need for a connected communication topology—an assumption that, although common, might be restrictive. More in detail, the observer design problem is divided into two subproblems to address separately the boundedness of the error dynamics in the detectable and undetectable subspaces. Eventually, LMI conditions are derived to keep the errors bounded and minimize the impact of measurement noise on its evolution.

The paper is organized as follows: Section \ref{sec:problem_formulation} outlines the distributed estimation problem and the distributed unknown input observer design problem. Section \ref{sec:structure_of_DUIO} presents the algorithm underlying the distributed observer. Section \ref{sec:design_of_DUIO} describes the design process and provides formal guarantees for the observer’s stability. Section \ref{sec:numerical_example} demonstrates the performance of the proposed strategy through two illustrative examples. Finally, Section \ref{sec:conclusions} offers concluding remarks and suggests directions for future research.
%\clearpage
%%%%%%%%%%%%%%%%%%%%%
%%%%%%%%%%%%%%%%%%%%%
%%%%%%%%%%%%%%%%%%%%%
\section{Notation}
\label{notation}
The symbols $\mathbb{N}$ and $\mathbb{R}$ denote the set of Natural and Real numbers, respectively. The matrix $I_n$ denotes the identity matrix of dimension $\mathbb{R}^{n \times n}$, while $\mathbf{0}_{n \times m}$ represents the zero matrix of dimension $\mathbb{R}^{n \times m}$. The operators $\operatorname{blkdiag}(\cdot)$ and $\operatorname{diag}(\cdot)$ are used to construct a block diagonal matrix and a diagonal matrix, respectively. Similarly, the operator $\operatorname{col}(\cdot)$ is used to construct a column vector.  
The operators $\operatorname{rank}(\cdot)$, $\operatorname{im}(\cdot)$, and $\operatorname{ker}(\cdot)$ represent the rank, image, and kernel of a matrix, respectively. Given a subspace $\mathcal{S} \in \mathbb{R}^n$ with dimension $\dim \mathcal{S}$ it's orthogonal complement is $\mathcal{S}^\perp$.
The symbol $\otimes$ represents the Kronecker product, while $\|\cdot\|_2$ denotes the Euclidean (or spectral) norm. 

%%%%%%%%%%%%%%%%%%%%%
%%%%%%%%%%%%%%%%%%%%%
%%%%%%%%%%%%%%%%%%%%%
\section{Problem Formulation}
\label{sec:problem_formulation}
Consider the discrete-time linear time-invariant system
\begin{equation}
    x(k + 1) = A x(k) + B u(k) + B_{w} w(k)
    \label{eq:d_system}
\end{equation}
where $k \in \mathbb{N}$ is the discrete time, $x(k) \in \mathbb{R}^{n_x}$ is the system state, $u(k) \in \mathbb{R}^{n_u}$ is the system input and $w(k) \in \mathbb{R}^{n_{{w}}}$ is an unknown disturbance. The matrices $A \in \mathbb{R}^{n_x \times n_x}$, $B \in \mathbb{R}^{n_x \times n_u}$ and $B_w \in \mathbb{R}^{n_x \times n_w}$ are the state transition matrix, the system input matrix and the disturbance gain matrix respectively. A network of $m$ sensors in monitoring system (\ref{eq:d_system}) where the output of each sensor is 
\begin{equation}
    y_i(k) = C_i x(k) + v_i(k), \quad \forall i = 1,\dots,m
\end{equation}
where $y_i \in \mathbb{R}^{n_{iy}}$, $C_i \in \mathbb{R}^{n_{iy}\times n_x}$ is the output matrix and $v_i(k) \in \mathbb{R}^{n_{iy}}$ is measurement noise for sensor $i$. Sensors are associated to nodes where local computation and data exchange take place. Nodes can communicate only with their immediate neighbors according to an underlying communication topology modeled as an undirected graph $\mathcal{G \coloneqq \left(\mathcal{N}, \mathcal{E}, \mathcal{A}_g\right)}$ where $\mathcal{N} \coloneqq \{1, \dots, m\}$ is a finite nonempty node set, $\mathcal{E} \subseteq \mathcal{N} \times \mathcal{N}$ is the set of edges and $\mathcal{A}_g \coloneqq \left[\alpha_{ij}\right]$ is the adjacency matrix. 
For each node $i$ its neighboring set $\mathcal{N}_i$ is defined as the set of all agents $j$ such that $(i, j) \in \mathcal{E}$, i.e. $\mathcal{N}_i \coloneqq \{j \,|\, (i, j) \in \mathcal{E}\}$.
%For each $(i,j) \in \mathcal{E}$ the corresponding entry of the adjacency matrix is $\alpha_{ij} > 0$ and $\alpha_{ij} = 0$ otherwise. 

\begin{figure}[ht!]
    \vspace{-1mm}
    \centering
    \includegraphics[width=\linewidth]{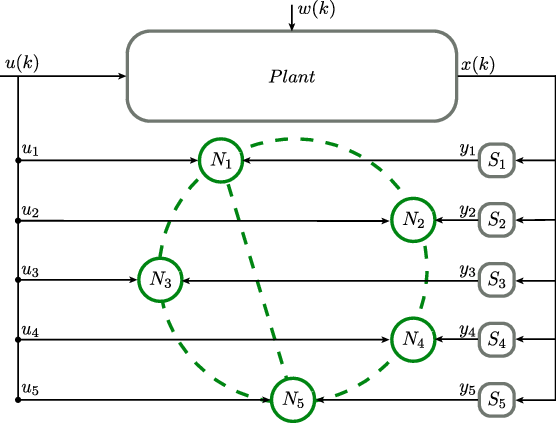}
    \caption{Distributed Observer Scheme: Each node $N_i$ collects locally available data, including the sensor output $y_i$ from its corresponding sensor $S_i$ and the local input $u_i$.}
    \label{fig:general_setup}
    \vspace{-1mm}
\end{figure}

The Laplacian matrix of $\mathcal{G}$ is $\mathcal{L}_g = \mathcal{D}_g - \mathcal{A}_g = [l_{ij}]$ where $\mathcal{D}_g \coloneqq diag([d_{1}, d_2, \dots, d_m]^T)$ is the degree matrix defined as the diagonal matrix in which each diagonal term $d_i$ for $i = 1,\dots, m$ identifies the number of edges connected to node $i$.
Sensor outputs $y_i(k)$ can be stacked into the following aggregated output
\begin{equation}
    y(k) = C x(k) + v(k)
\end{equation}
The aggregated output matrix and measurement error are 
\begin{equation}
    \hspace{-2mm}C = \begin{bmatrix}
        C_1^T & \dots C_m^T
    \end{bmatrix}^T, v(k) = \begin{bmatrix}
        v_1(k)^T & \dots v_m(k)^T
    \end{bmatrix}^T
\end{equation}
% where the vector $v(k)$ is a zero-mean independent and identically distributed (i.i.d) Gaussian noise with covariance $Q_v \geq 0$.
To clearly differentiate locally available signals, we partition the system inputs into two distinct components: a component $u_i$, known at node $i$, and a component $u_i^u$, which is unknown to $i$ and can instead be seen as an exogenous disturbance, i.e. 
\begin{equation}
    Bu(k) = B_iu_i(k) + B^u_iu^u_i(k)
\end{equation}
where $u_i \in \mathbb{R}^{n_{iu}}$, $B_i \in \mathbb{R}^{n_x \times n_{iu}}$, $u_i^u \in \mathbb{R}^{n_{iu}^u}$, $B^u_i \in \mathbb{R}^{n_x \times n_{iu}^u}$ and $n_u = n_{iu} + n_{iu}^u$.
As a consequence, the augmented local unknown disturbance vector is defined as $ \bar w_i(k) := [(u^u_i(k))^T, w(k)^T]^T$ while the augmented local unknown input matrix is denoted with
$\bar B_{iw} := [B_i^u, B_w]$. A graphical representation of the described setup is given in Figure \ref{fig:general_setup}.

Throughout this work, we assume that the following conditions hold:
\begin{assumption}
    $\operatorname{rank}(\bar B_{iw}) = n_w + n^u_{iu}$ and $\operatorname{rank}(C_i) = n_{iy}$ for all $i\in \mathcal{N}$.
    \label{ass:rank_B}
\end{assumption}
\begin{assumption}
    $\operatorname{rank}(C_i \bar B_{iw}) = n_w + n^u_{iu}$ for all $i\in \mathcal{N}$.
    \label{ass:rank_CB}
\end{assumption}
\begin{assumption}
    The measurement noise $v_i(k)$ is bounded for all $k \geq 0$ and $\forall i\in \mathcal{N}$.
    \label{ass:bounded_meas_err}
\end{assumption}
\begin{remark}
    There is no loss of generality in assuming that the local matrix $\bar B_{iw}$ and the local output matrix $C_i$ have full column rank and full row rank, respectively \citep{CPZ:96}.  In addition, the rank condition about their product is commonly used for the classic centralized UIO design \citep{YBRP:22,TTF:08}.
\end{remark}
The problem addressed in this work can be stated as follows:
\begin{problem}
    (\textbf{Distributed UIO Design}) \textit{Given the system (\ref{eq:d_system}) under Assumptions \ref{ass:rank_B}, \ref{ass:rank_CB}, and \ref{ass:bounded_meas_err}, design a distributed unknown input observer (D-UIO) that, for each node $i = 1, \dots, m$, reconstructs a complete estimate $\hat{x}_i(k)$ of the internal state $x(k)$ using only local and neighboring data, even in the presence of the unknown disturbance $\bar{w}_i(k)$ and the measurement error $v_i(k)$.}
    \label{prob:main}
\end{problem}
\section{Structure of the Distributed Unknown Input Observer}
\label{sec:structure_of_DUIO}
Solving Problem \ref{prob:main} requires achieving two objectives: (i) establishing rules for the distributed estimation of the state $x(k)$, and (ii) devising a strategy to locally mitigate the effects of both disturbances $\bar{w}_i(k)$ and measurement noise $v_i(k)$. To address both objectives simultaneously, we propose a distributed Unknown Input Observer. The design of the UIO is based on decomposing the state space into locally detectable and undetectable components and leveraging some algebraic properties of invariant subspaces. The following sections outline the rank conditions required for the UIO design and detail the \textit{detectability decomposition} that is at the base of the proposed design process.

\subsection{Rank condition for UIO existence}
To decouple the UIO dynamics from the unknown input $\bar{w}_i(k)$ at each node $i \in \mathcal{N}$, the following conditions must hold
\begin{equation}
    P_i\bar B_{iw} = 0, \quad \forall i\in \mathcal{N}
    \label{eq:rank_con}
\end{equation}
where $P_i \in \mathbb{R}^{n_x\times n_x}$ is a matrix defined as follows
\begin{equation}
    P_i \coloneqq  I_{n_x} + E_iC_i
    \label{eq:pi_definition}
\end{equation}
and $E_i \in \mathbb{R}^{n_x\times n_{iy}}$ is matrix to be designed. It can be proved that $E_i$ solving \eqref{eq:rank_con}  exists if and only if Assumption \ref{ass:rank_CB} holds \citep{CPR:12}. 
The general solution of (\ref{eq:rank_con}) can be expressed as  
\begin{equation}
    E_i = -\bar{B}_{iw}(C_i\bar{B}_{iw})^\dagger + H_i(I - C_i\bar{B}_{iw}(C_i\bar{B}_{iw})^\dagger)
    \label{eq:e_i_definition}
\end{equation}
where $(C_i\bar{B}_{iw})^\dagger$ denotes the left inverse of $C_i\bar{B}_{iw}$, given by  
\begin{equation}
    (C_i\bar{B}_{iw})^\dagger = \left((C_i\bar{B}_{iw})^T(C_i\bar{B}_{iw})\right)^{-1}(C_i\bar{B}_{iw})^T
\end{equation}
and $H_i$ is an arbitrary matrix of appropriate dimensions.  
A well-chosen $E_i$ maximizes the rank of $P_i$, ensuring that  $\ker(P_i) = \operatorname{im}(\bar{B}_{iw})$ which leads to the condition  
\begin{equation}
    \operatorname{rank} \begin{bmatrix}
        P_i\\
        \bar{B}_{iw}^\dagger
    \end{bmatrix} = n_x.
    \label{eq:rank_con2}
\end{equation}
In general, the most common choice for $E_i$ to satisfy the rank condition (\ref{eq:rank_con2}) 
is  $E_i = -\bar{B}_{iw} (C_i \bar{B}_{iw})^\dagger$.

\subsection{Detectability decomposition}
The key idea behind the design of the distributed unknown input observer is to identify a similarity transformation that, when applied to the matrix $P_i A$, induces a triangular structure. 
%To achieve this, invariant subspaces play a crucial role. 
%As is well known, the unobservable subspace of the pair $(P_iA, C_i)$ is the largest $P_iA$-invariant subspace contained within the kernel of $C_i$ \citep{W:74}.  In \citep{WLMA:19}, an observability decomposition is introduced determine observable and unobservable subspaces for each node $i$, thereby achieving the desired triangular structure. In this work, to obtain a more general formulation, we employ a \textit{detectability decomposition} instead. This approach partitions the state space into \textit{detectable} and \textit{undetectable} subspaces, 
To this end, the detectability decomposition theory \citep{CD:12} is used to determine the detectable and undetectable subspaces of the pair of matrices $(P_iA, C_i)$ at node $i$ denoted as $\mathcal{D}_i$ and $\mathcal{U}_i$, respectively. Let $\chi_{P_iA}(z)$ denote the minimal polynomial of $P_iA$, defined as the monic polynomial of the lowest degree satisfying  $\chi_{P_iA}(P_iA) = 0$. The coprime factorization $\chi_{P_iA}(z) = \chi_{P_iA}^-(z) \chi^+_{P_iA}(z)$
corresponds to a decomposition of the space $\mathbb{R}^{n_x}$, where $\ker(\chi_{P_iA}^-(P_iA))$ contains the states with strictly stable dynamics and $\ker(\chi_{P_iA}^+(P_iA))$ contains states with marginally stable or unstable dynamics. Consequently, the undetectable and  detectable subspaces of $(P_iA, C_i)$ are defined as:
\begin{gather}
    \mathcal{U}_i \coloneqq  \bigcap_{l=1}^{n_x} \mbox{Ker}\left(C_i (P_iA)^{l-1}\right)  \cap \operatorname{Ker} \left(\chi_{P_iA}^+(P_iA)\right) \\
    \mathcal{D}_i \coloneqq \mathcal{U}_i^\perp
\end{gather}
where $\dim \mathcal{U}_i = \nu_i$ and $\dim \mathcal{D}_i = n_x - \nu_i$. To construct the similarity transformation, we introduce two isometries, $D_i \in \mathbb{R}^{n_x \times (n_x - \nu_i)}$ and $U_i \in \mathbb{R}^{n_x \times \nu_i}$, satisfying  $D_i^T D_i = I_{n_x - \nu_i}, \quad U_i^T U_i = I_{\nu_i}$,
where the column spaces of $D_i$ and $U_i$ are $\mathcal{D}_i$ and $\mathcal{U}_i$, respectively.
Noting also that
\begin{equation}
    \operatorname{Im} D_i =\operatorname{Im}\left(
    \begin{bmatrix}
        C_i\\
        C_i P_i A\\
        \vdots\\
        C_i \left(P_iA\right)^{n_x - 1}\\
        \chi_{P_iA}^+(P_iA)
    \end{bmatrix}^T\right)
\end{equation}
 $D_i$ can be obtained using the Gram–Schmidt procedure, while $U_i$ is chosen such that  $[D_i \quad U_i]$ forms an orthonormal matrix. 
As a result, the detectability decomposition for matrices $P_iA$ and $C_i$ of system (\ref{eq:d_system}) is given by  
\begin{gather}
    \begin{bmatrix}
        D_i^T\\
        U_i^T
    \end{bmatrix}
    P_i A
    \begin{bmatrix}
        D_i & U_i
    \end{bmatrix}
    =
    \begin{bmatrix}
        A_{id} & 0\\
        A_{ir} & A_{iu}
    \end{bmatrix}\label{eq:decomposition1}\\
    C_i
    \begin{bmatrix}
        D_i & U_i
    \end{bmatrix} =
    \begin{bmatrix}
        C_{id} & 0
    \end{bmatrix}
    \label{eq:decomposition2}
\end{gather}
where $A_{id} \coloneqq D_i^TP_iAD_i \in \mathbb{R}^{(n_x - \nu_i) \times (n_x - \nu_i)}$, $A_{ir} \coloneqq D_i^TP_iAU_i \in \mathbb{R}^{\nu_i \times (n_x - \nu_i)}$, $A_{iu} \coloneqq U_i^TP_iAU_i \in \mathbb{R}^{\nu_i \times \nu_i}$ and $ C_{id} \coloneqq C_i D_i \in \mathbb{R}^{n_{iy}\times(n_x - \nu_i)}$. Similarly to a Kalman canonical representation, the pair $(A_{id}, C_{id})$ is observable while, because of the detectability decomposition, the matrix $A_{iu}$ is unstable.

\subsection{Distributed Unknown Input Observer Algorithm}

% \begin{figure}[hb]
% {
% \small
% \noindent\rule{\columnwidth}{0.05cm}
% {\textbf{Algorithm 1: Distributed Unknown Input Observer} }\\
% \noindent\!\!\rule{\columnwidth}{0.05cm}
% \begin{algorithmic}[1]
% \vspace{-4mm}
% \STATE {
% Collect state estimates $\hat x_j(k)$ $\forall j\in \mathcal{N}_i$ from neighboring nodes
% }
% \STATE {
% Reconstruct local estimate as
% \begin{gather*}
% \hat x_i(k) = \xi_i(k) - E_i y_i(k) 
% \end{gather*}
% }
% \STATE {
% Update the internal node state as
% \begin{gather*}
% \begin{multlined}
%     \xi_i(k + 1) = (P_i A - K_i C_i) \xi_i(k) + L_i y_i(k) \\ + P_iB_i u_i(k) - G_i \sum_{j = 1}^m \alpha_{ij}\left(\hat x_i(k) - \hat x_j(k)\right)
% \end{multlined}
% \end{gather*}
% }
% \STATE{
% Transmit the reconstructed state $\hat x_i(k)$ to neighbors
% }
% \end{algorithmic}
% \rule{\columnwidth}{0.05cm}
% where $\xi_i(0) = \mathbf{0}_{n_x}$, $\hat x_i(0) = \mathbf{0}_{n_x}$ for all $i = 1, \dots, m$.
% }
% \end{figure}

The UIO is characterized by the following update rules at each node $i$:
%\begin{gather}
%\begin{multlined}
\begin{align}
    \xi_i(k + 1) =& (P_i A - K_i C_i) \xi_i(k) + L_i y_i(k) 
    + P_iB_i u_i(k)\nonumber \\ 
    &- G_i \sum_{j = 1}^m \alpha_{ij} \left(\hat{x}_i(k) - \hat{x}_j(k)\right)\label{eq:uio_structure_z}\\
\hat{x}_i(k) =& \xi_i(k) - E_i y_i(k) \label{eq:uio_structure_hatx}
\end{align}
%\end{multlined}\label{eq:uio_structure_z}\\
where $\xi_i(k) \in \mathbb{R}^{n_x}$ represents the internal state of node $i$, and $\hat{x}_i(k)$ is the estimate of the system state $x(k)$ computed at node $i$ at time step $k$.
 To eliminate the dependence of the observer evolution on the unknown input, the matrix $E_i$ is computed as in (\ref{eq:e_i_definition}), setting $H_i = 0_{n_x \times n_{iy}}$. The matrix $L_i$ is then defined as  
\begin{equation}
    L_i \coloneqq K_i(I + C_iE_i) - (I + E_iC_i)AE_i
    \label{eq:li_definition}
\end{equation}
while gains $K_i$ and $G_i$ are defined as follows:  
\begin{gather}
    K_i \coloneqq D_i^T K_{id}, \quad G_i \coloneqq g_i U_i U_i^T
    \label{eq:gains}
\end{gather}
where $K_{id} \in \mathbb{R}^{n_x \times n_{iy}}$ and $g_i \in \mathbb{R}$ are design parameters. As is going to be clarified later, this particular structure for the gains $K_i$ and $G_i$ enable to decouple the design of the two gains. The Distributed Unknown Input Observer strategy is summarized in the following Algorithm where the initialization values are $\xi_i(0) = \mathbf{0}_{n_x}$, for all $i = 1, \dots, m$.

\begin{algorithm}
    \caption{Distributed Unknown Input Observer}\label{alg:DUIO}
    \begin{algorithmic}[1]
    \State {
    Collect state estimates $\hat x_j(k)$ $\forall j\in \mathcal{N}_i$ from neighboring nodes
    }
    \State { Reconstruct local estimate as
    \begin{gather*}
    \hat x_i(k) = \xi_i(k) - E_i y_i(k) 
    \end{gather*}
    }
    \State {
    Update the internal node state as
    \begin{gather*}
    \begin{multlined}
        \xi_i(k + 1) = (P_i A - K_i C_i) \xi_i(k) + L_i y_i(k) \\ + P_iB_i u_i(k) - G_i \sum_{j = 1}^m \alpha_{ij}\left(\hat x_i(k) - \hat x_j(k)\right)
    \end{multlined}
    \end{gather*}
    }
    \State Transmit the reconstructed state $\hat x_i(k)$ to neighbors
\end{algorithmic}
\end{algorithm}

\begin{remark}
 When $B_w = 0$ i.e., the system is not affected by unknown inputs, and all nodes have full access to input data the trivial solution $E_i = 0$, $\forall i \in \mathcal{N}$, can be adopted, reducing the proposed D-UIO to the discrete-time counterpart of the distributed Luenberger observer proposed in \citep{KSC:16}.
\end{remark}

\section{Design of the D-UIO parameters}
\label{sec:design_of_DUIO}
This section details the procedure to calculate the design parameters $K_{id}$ and $g_i$ for $i = 1,\dots, m$.
\subsection{Analyses of the error evolution}
Let us define the state reconstruction error for the $i$-th observer as $e_i(k) = \hat{x}_i(k) - x(k)$, and its concatenated form as $e(k) = \operatorname{col}(e_i)_{i \in \mathcal{N}}$. To facilitate the analysis of the error dynamics let us first analyze the term $L_i v_i(k)$. In particular, its contribution in the detectable subspace is given by
\begin{align}
D_i^T L_i v_i(k) =& D_i^T\left[K_i(I + C_i E_i)\right. \nonumber\\
&\left.- (I + E_i C_i)A E_i\right] v_i(k) \nonumber \\
% &= D_i^T K_i v_i(k) + D_i^T K_i C_i E_i v_i(k) - D_i^T P_i A E_i v_i(k) \nonumber \\
=& D_i^T K_i v_i(k) \nonumber\\
&+ D_i^T K_i C_i (D_i D_i^T + U_i U_i^T) E_i v_i(k) \nonumber\\
& - D_i^T P_i A (D_i D_i^T + U_i U_i^T) E_i v_i(k)  \label{eq:dlv_step1}
\end{align}
where we have used the orthogonality of the matrix $[D_i \quad U_i]$, in particular the identity $D_i D_i^T + U_i U_i^T = I_{n_x}$, as well as the definitions of the matrices $P_i$ and $L_i$ provided in (\ref{eq:pi_definition}) and (\ref{eq:li_definition}). Substituting (\ref{eq:gains}) and (\ref{eq:decomposition1})-(\ref{eq:decomposition2}) into (\ref{eq:dlv_step1}), we obtain
\begin{align}
    D_i^T L_i v_i(k) =& K_{id} v_i(k) + K_{id} C_{id} E_i v_i(k) - A_{id} E_i v_i(k) \nonumber \\
    =& -(A_{id} - K_{id} C_{id}) D_i^T E_i v_i(k) + K_{id} v_i(k)
    \label{eq:dlv_steps2}
\end{align}
Following the same methodology, in the undetectable subspace we have
\begin{align}
U_i^T L_i v_i(k) &= U_i^T\left[K_i(I + C_i E_i) - (I + E_i C_i)A E_i\right] v_i(k) \nonumber \\
&= - U_i^T P_i A (D_i D_i^T + U_i U_i^T) E_i v_i(k) \nonumber \\
&= - A_{ir} D_i^T E_i v_i(k) - A_{iu} U_i^T E_i v_i(k)
\end{align}
where, in the second equation, we have used the identity $U_i K_i = U_i D_i^T K_{id} = 0$, which follows from the orthogonality condition $U_i^T D_i = 0$.
Before proceeding to the error analysis, let us also examine the term $L_i y_i(k)$. In particular, within the detectable subspace, we have
\begin{align}
    D_i^TL_iy_i(k) =& D_i^TL_i(C_i x(k) + v_i(k))\nonumber\\
        =&  D_i^TL_iC_i x(k) + D_i^T L_iv_i(k)\nonumber\\
        % =&D_i^T(K_i(I + C_iE_i) -  \nonumber\\
        % & (I + E_iC_i)AE_i)C_ix(k) + D_i^T L_iv_i(k)\nonumber\\
        =& K_{id}C_ix(k) + K_{id}C_iE_i - D_i^TAE_iC_ix(k) \nonumber\\
         &- D_i^TE_iC_iAE_iC_ix(k) + D_i^T L_i v_i(k)\nonumber\\
        % =&K_{id}C_i(I + C_iE_i)x(k) -\nonumber\\
        % & D_i^T(I - E_iC_i)AE_iC_ix(k) + D_i^TL_iv_i(k)\nonumber\\
        =& K_{id}C_i(D_iD_i^T + U_iU_i^T + E_iC_i)x(k) -\nonumber\\
         &D_i^T(I + E_iC_i)A(D_iD_i^T+U_iU_i^T)E_iC_ix(k) \nonumber\\
         &+ D_i L_iv_i(k)\nonumber\\
        % =&K_{id}C_i(D_iD_i^T + E_iC_i)x(k) - \nonumber\\
        % &D_i^T(I+E_iC_i)AD_iD_i^TE_iC_ix(k) + \nonumber\\
        % &D_i^TL_iv_i(k)\nonumber\\
        =& -(A_{id} - K_{id}C_i)D_i^TE_iy_i(k) \nonumber\\
        &+ K_{id}C_{id}D_i^Tx(k) + K_{id}v_i(k)
\end{align}
On the other hand, in the undetectable subspace, we obtain
\begin{align}
    U_i^TL_iy_i(k) =& U_i^TL_i(C_i x(k) + v_i(k))\nonumber\\
    =&  U_i^TL_iC_i x(k) + U_i^T L_i v_i(k)\nonumber\\
    =&  U_i^T(K_i(I + C_iE_i) \nonumber\\
    &- (I + E_iC_i)AE_i)C_ix(k)\nonumber\\
     &+ U_i^TL_i v_i(k)\nonumber\\
    %
    % =& U_i^TK_i(I + C_iE_i)C_ix(k) - \nonumber\\ 
    % & U_i^T(I+C_iE_i)AE_iC_ix(k) + U^TL_iv(k)\nonumber\\
    %
   =& U_i^TL_iv(k) -
    U_i^T(I + E_iC_i)A(D_iD_i^T \nonumber\\
    &+ U_iU_i^T)E_iC_ix(k)\nonumber\\
    =& - (A_{ir}D_i^T + A_{iu}U_i^T)E_iy(k)
    \label{eq:ulv}
\end{align}
We can now proceed with the analysis of the error evolution. For node $i$, let us define the error evolution in the detectable subspace as $e_{id}^k \coloneqq D_i^T e_i(k)$. Consequently, we have
\begin{align}
    e_{id}^{k + 1} =& D_i^T(\hat x_i(k + 1)-  x(k + 1))\nonumber\\
    =& D_i^T( \xi_i(k + 1) - E_iy_i(k + 1) - x(k + 1))\nonumber\\
    %
    % =&D_i^T((P_i A - K_i C_i)\xi_i(k) + L_i y_i(k) \nonumber\\
    % & + D_i^TP_iB_iu_i(k) - G_i \sum_{j = 1}^m \alpha_{ij} \left( \hat x_i(k) - \hat x_j(k)\right) \nonumber\\
    % & - E_iy_i(k + 1) - x(k + 1))\nonumber\\
    %
    =& D_i^T(P_i A - K_i C_i)\xi_i(k) + D_i^T L_i y_i(k) \nonumber\\
    & + D_i^TP_iB_iu_i(k) - D_i^T G_i \sum_{j = 1}^m \alpha_{ij} \left( \hat x_i(k) - \hat x_j(k)\right) \nonumber\\
    & - D_i^T E_iy_i(k + 1) - D_i^T x(k + 1)\nonumber\\
    %
    % =& D_i^T(P_i A - K_i C_i)\xi_i(k) + D_i^T L_i y_i(k)\nonumber\\
    % &+ D_i^TP_iB_iu_i(k) - D_i^T E_iy_i(k + 1) - D_i^T x(k + 1)\nonumber\\
    % %
    % =& D_i^T(P_i A - K_i C_i)\xi_i(k) + D_i^T L_i y_i(k)\nonumber\\
    % &+ D_i^TP_iB_iu_i(k) - D_i^T E_i C_i x(k + 1) - D_i^T x(k + 1)\nonumber\\
    % & - D_i^T E_i v_i(k + 1)\nonumber\\
    % %
    =& D_i^T(P_i A - K_i C_i)\xi_i(k) + D_i^T L_i y_i(k) \nonumber\\
    &- D_i^T E_i v_i(k + 1) + D_i^TP_iB_iu_i(k) \nonumber\\
    &- D_i^T P_ix(k + 1)
    \label{eq:e_id_step1}
\end{align}
where we have used the update rules given in (\ref{eq:uio_structure_hatx}) and the identity $D_i^T G_i = 0$, which follows directly from the orthogonality condition $D_i^T U_i = 0$. Substituting (\ref{eq:d_system}) into (\ref{eq:e_id_step1}) and performing some algebraic manipulations, we obtain
\begin{align}
    e_{id}^{k+1}
    % %
    =& D_i^T(P_i A - K_i C_i)\xi_i(k) + D_i^T L_i y_i(k)\nonumber\\
    &+ D_i^TP_iB_iu_i(k) - D_i^TP_iAx(k) - D_i^TP_iB_iu_i(k) \nonumber\\
    &- D_i^TP_i\bar B_{iw} \bar w_i(k) - D_i^T E_i v_i(k + 1)\nonumber\\
    % %
    =& D_i^T(P_i A - K_i C_i)(D_iD_i^T + U_iU_i^T)\xi_i(k) \nonumber\\
    & - D_i^TP_iA(D_iD_i^T + U_iU_i^T)x(k) \nonumber\\  
    & + D_i^T L_i y_i(k) - D_i^T E_i v_i(k + 1)\nonumber\\
    % %
     %
    % =& D_i^T P_i A D_iD_i^T\xi_i(k) - D_i^T K_i C_i D_iD_i^T\xi_i(k)\nonumber\\
    % &+ D_i^T P_i A U_iU_i^T\xi_i(k) - D_i^T K_i C_i U_iU_i^T\xi_i(k) \nonumber\\
    % & -D_i^T P_i A D_iD_i^Tx(k) - D_i^T P_i A U_iU_i^Tx(k) \nonumber\\
    % &+ D_i^T L_i y_i(k) - D_i^T E_i v_i(k + 1)\nonumber\\
    % %
    =& A_{id} D_i^T\xi_i(k) - D_i^T D_i K_{id} C_{id}D_i^T\xi_i(k) \nonumber \\
    & -A_{id} D_i^Tx(k) + D_i^T L_i y_i(k) - D_i^T E_i v_i(k + 1) \nonumber\\
    % %
    =& (A_{id} - K_{id} C_{id})D_i^T\xi_i(k) -A_{id} D_i^Tx(k) \nonumber \\
    & + D_i^T L_i y_i(k) - D_i^T E_i v_i(k + 1)\nonumber\\
    % %
    =& (A_{id} - K_{id} C_{id})D_i^T(\xi_i(k) -E_iy(k) - x(k)) \nonumber \\
    &  + K_{id}v_i(k) - D_i^T E_i v_i(k + 1)
    \label{eq:eid_step2}
\end{align}
where the second equality leverages the selection of $E_i$ satisfying condition (\ref{eq:rank_con}), effectively decoupling the error evolution from the unknown input $\bar w(k)$. Similarly, let us define $e_{iu}^{k} \coloneqq U_i^T e(k)$. Following the same steps as in (\ref{eq:e_id_step1})-(\ref{eq:eid_step2}), the error dynamics in the undetectable subspace can be expressed as
\begin{align}
    e_{iu}^{k+1} =& U_i^T(\hat x_i(k + 1) -  x(k + 1))\nonumber\\
    =& U_i^T( \xi_i(k + 1) - E_iy_i(k + 1) - x(k + 1))\nonumber\\
    %
    % =&U_i^T((P_i A - K_i C_i)\xi_i(k) + L_i y_i(k) \nonumber\\
    % & + P_iB_iu_i(k) - G_i \sum_{j = 1}^m \alpha_{ij} \left( \hat x_i(k) - \hat x_j(k)\right) \nonumber\\
    % & - E_iy_i(k + 1) - x(k + 1))\nonumber\\
    %
    =&U_i^T(P_i A - K_i C_i)\xi_i(k) + U_i^TL_i y_i(k) \nonumber\\
    & + U_i^TP_iB_iu_i(k) - g_iU_i^T \sum_{j = 1}^m \alpha_{ij} \left( e_i(k) - e_j(k)\right) \nonumber\\
    & - U_i^TE_iy_i(k + 1) - U_i^Tx(k + 1)\nonumber\\
    % 
    % =&U_i^T(P_i A - K_i C_i)\xi_i(k) + U_i^TL_i y_i(k) \nonumber\\
    % & + U_i^TP_iB_iu_i(k) - g_iU_i^T \sum_{j = 1}^m \alpha_{ij} \left( e_i(k) - e_j(k)\right) \nonumber\\
    % & - U_i^TP_ix(k + 1) - U_i^TE_iv_i(k + 1)\nonumber\\
    % 
    =&U_i^T(P_i A - K_i C_i)\xi_i(k) + U_i^TL_i y_i(k) \nonumber\\
    & + U_i^TP_iB_iu_i(k) - g_iU_i^T \sum_{j = 1}^m l_{ij} e_j(k) \nonumber\\
    & - U_i^TP_iAx(k) - U_i^TP_iBu(k) - U_i^TP_i\bar B_{iw}\bar w_i(k) \nonumber\\
    &- U_i^TE_iv_i(k + 1)
    \label{eq:eiu_step1}
\end{align}
Noting that the summation in (\ref{eq:eiu_step1}) can be rewritten as $\sum_{j = 1}^m \alpha_{ij} \left( e_i(k) - e_j(k)\right) = \sum_{j = 1}^m l_{ij} e_j(k)$, after some algebraic rearrangements we obtain
\begin{align}
     e_{iu}^{k+1}
    =&U_i^T(P_i A - K_i C_i)(D_iD_i^T + U_iU_i^T)\xi_i(k) \nonumber\\
    & - g_iU_i^T \sum_{j = 1}^m l_{ij} (D_jD_j^T + U_jU_j^T)e_j(k) \nonumber\\
    & - U_i^TP_iA(D_iD_i^T + U_iU_i^T)x(k) \nonumber\\
    &+ U_i^TL_i y_i(k) - U_i^TE_iv_i(k + 1) \nonumber\\
    =&A_{ir}D_i^T\xi_i(k) + A_{iu}U_i^T\xi_i(k) + U_i^TL_i y_i(k) \nonumber\\
    & - g_iU_i^T \sum_{j = 1}^m l_{ij} D_jD_j^T e_j(k) \nonumber\\
    & - g_iU_i^T \sum_{j = 1}^m l_{ij} U_jU_j^T e_j(k) \nonumber\\
    & -A_{ir}D_i^Tx(k) - A_{iu}U_i^Tx(k)  - U_i^TE_iv_i(k + 1)\nonumber\\
    =&A_{ir}D_i^Te_i(k)  - g_iU_i^T \sum_{j = 1}^m l_{ij} D_jD_j^T e_j(k) \nonumber\\
    & + A_{iu}U_i^Te_i(k) - g_iU_i^T \sum_{j = 1}^m l_{ij} U_jU_j^T e_j(k) \nonumber\\
    &- U_i^TE_iv_i(k + 1)
    \label{eq:error_usubspace}
\end{align}
To facilitate the analysis of the overall error dynamics, it is convenient to express the error evolution in the detectable and undetectable subspaces in aggregated form. The aggregated form of the error dynamics in the detectable subspace is given by
\begin{multline}
    %\begin{aligned}
        D^T e(k + 1) = (A_d - K_d C_d) D^T e(k)\\
       + K_d v(k) - D^T E v(k + 1)
    %\end{aligned}
    \label{eq:aggr_err_d}
\end{multline}
where $D = \operatorname{blkdiag}(D_i)_{i\in\mathcal{N}}$, $A_d = \operatorname{blkdiag}(A_{id})_{i\in\mathcal{N}}$, $C_d = \operatorname{blkdiag}(C_{id})_{i\in\mathcal{N}}$, $E = \operatorname{blkdiag}(E_i)_{i\in\mathcal{N}}$ and, $K_d = \operatorname{blkdiag}(K_{id})_{i\in\mathcal{N}}$.
On the other hand, the concatenated error dynamics in the undetectable subspace can be expressed as
\begin{equation}
    \begin{aligned}
        U^T e(k + 1) &= \left(A_u - \tilde{g} U^T (\mathcal{L}_g \otimes I_{n_x}) U\right) U^T e(k) \\
        &\quad + \left(A_r - \tilde{g} U^T (\mathcal{L}_g \otimes I_{n_x}) D\right) D^T e(k) \\
        &\quad - U^T E v(k + 1)
    \end{aligned}
    \label{eq:aggr_err_u}
\end{equation}
where $A_u = \operatorname{blkdiag}(A_{iu})_{i\in\mathcal{N}}$, $A_r = \operatorname{blkdiag}(A_{ir})_{i\in\mathcal{N}}$ and, $\tilde{g} = \operatorname{diag}(g_i I_{\nu_i})_{i\in \mathcal{N}}$.
Combining (\ref{eq:aggr_err_d}) and (\ref{eq:aggr_err_u}), the overall closed-loop error dynamics can be expressed as
\begin{equation}
\begin{aligned}
    e(k + 1) =&
    \begin{bmatrix}
        D & U
    \end{bmatrix}
    \begin{bmatrix}
        \Phi_d & 0\\
        \Phi_r & \Phi_u
    \end{bmatrix}
    \begin{bmatrix}
        D^T\\
        U^T
    \end{bmatrix}e(k)\\
    &+
    \begin{bmatrix}
        D & U
    \end{bmatrix}
    \begin{bmatrix}
        K_d\\
        0
    \end{bmatrix} v(k)
    - E v(k  + 1)
\end{aligned}
\label{eq:aggr_error}
\end{equation}
where $\Phi_d \coloneqq  A_d - K_dC_d$, $\Phi_r \coloneqq A_{r} -\tilde{g}U^T\left(\mathcal{L}_g\otimes I_{n_x}\right)D$, and $\Phi_u \coloneqq A_{u} - \tilde{g}U^T\left(\mathcal{L}_g\otimes I_{n_x}\right)U$.
It is important to underline that, due to the concealed lower-triangular structure of (\ref{eq:aggr_error}), the design of the D-UIO can be decoupled into two independent problems:
\begin{itemize}
    \item[(i)] The design of the gains $K_{id}$ for $i = 1, \dots, m$ that guarantee boundedness of the error dynamics in the locally detectable subspace;
    \item[(ii)] The design of the gains $g_i$ for $i = 1, \dots, m$ that guarantee boundedness of the error dynamics in the locally undetectable subspace.
\end{itemize}
In the following, the above two problems are addressed separately and the corresponding solutions are provided.

\subsection{Design of the stabilizing gain for the detectable subspace}
Given Assumption \ref{ass:bounded_meas_err}, the error evolution (\ref{eq:aggr_err_d}) (equivalently (\ref{eq:eid_step2}) for all $i \in \mathcal{N}$) remains bounded  if and only if the  closed-loop matrices $(A_{id} - K_{id}C_{id})$ are asymptotically stable for $i = 1,\dots, m$. Consequently, $K_{id}$ can be chosen such that the eigenvalues of $(A_{id} - K_{id}C_{id})$ lie within the unit circle. Notably, such a $K_{id}$ always exists because the pair $(A_{id}, C_{id})$ is Observable.

In this work, we exploit this degree of freedom by designing $K_{id}$ in such a way to mitigate the impact of the local measurement error $v_i(k)$. Specifically, we introduce the following performance variable
\begin{equation}
    z_i(k) = e_{id}(k), \quad \forall i \in \mathcal{N}
    \label{eq:z_index}
\end{equation}
% cerca di usare z
and decide to design $K_{id}$ by minimizing the following $H_\infty$ performance index
\begin{equation}
    J_i \coloneqq \max_{\tilde v_i \in \mathcal{V}_i} \frac{\|z_i(k)\|_2}{\|\tilde v_i(k)\|_2}
    \label{eq:h_inf_p}
\end{equation}
where $\mathcal{V}_i$ is the set of all possible realization of the overall bounded local disturbance $\tilde v_i(k) \coloneqq [v_i(k)^T, v_i(k + 1)^T]^T$.

% The error dynamic for the detectable sub-space  given in eq. (\ref{eq:error_dsubspace}) can be rearranged as
% \begin{equation}
% \begin{multlined}
%     e_{id}(k + 1) = (A_{id} - K_{id}C_{id})e_{id}(k) + \\\left(
%     \begin{bmatrix}
%         0 & -D_i^TE_i
%     \end{bmatrix}
%     + K_{id}
%     \begin{bmatrix}
%         I & 0
%     \end{bmatrix}\right)\tilde v_i(k).
% \end{multlined}
% \label{eq:extracted_ddynamic}
% \end{equation}
% where $\tilde v_i(k) \coloneqq [v_i(k)^T, v_i(k + 1)^T]^T$.
% Considering that each gain can be calculated separately we propose here to design $K_{id}$ for $i = 1,\dots, m$ in such a way to stabilize the dynamic (\ref{eq:extracted_ddynamic}) and minimize the $H_{\infty}$ norm with respect the following performance index:
% \begin{equation}
%     p(k) = e_{id}(k).
%     \label{eq:z_index}
% \end{equation}
\begin{proposition}
    Let Assumption \ref{ass:bounded_meas_err} hold true. Given the closed-loop error dynamics (\ref{eq:eid_step2}) and the $H_{\infty}$ performance index (\ref{eq:h_inf_p}), if  matrices $P^* = (P^*)^T > 0$, $Y_d^*$ and a scalar $\beta_{id}^*$ exist that solve the following semidefinite programming (SDP) optimization problem:  
    \begin{equation}
        \begin{aligned}
            &[P^*, Y_d^*, \beta_{id}^*] = \arg \min_{P, Y_d, \beta_{id}} \quad\beta_{id}\\
            &\text{s. t.} 
            \begin{bmatrix}
                P & P A_{id} - Y_d C_{id} & -[Y_d \,\,PD_i^TE_i] & 0\\
                * & P & 0 & I\\
                * & * & \beta_{id} I & 0\\
                * & * & * & \beta_{id} I
            \end{bmatrix} > 0\\
             &\quad\,\,\,\,\,P > 0
        \end{aligned}
        \label{eq:problem_kinf}
    \end{equation}  
then the gain $K_{id} = (P^*)^{-1}Y_d^*$ ensures that the closed-loop error dynamics (\ref{eq:eid_step2}) remains bounded, with a performance guarantee $J_i \leq \beta^*_{id}$.
\end{proposition}
\begin{proof}
Given the following generic closed-loop system
\begin{equation}
    \begin{aligned}
        s(k + 1) &= (A_{s} - K_{s}C_{1s})s(k) +(B_{s} - K_{s}D_{1s})\tilde v_i(k)\\
        z_{s}(k) &= C_{2s}s(k) + D_{2s} \tilde v_i(k),
    \end{aligned}
    \label{eq:generic_model}
\end{equation}
where $\tilde v_i(k)$ is a disturbance signal, the optimal $H_{\infty}$ gain $K_s$ with performance guarantee $\beta_{id}$ can be computed by solving the following optimization problem \citep{DY:13}:
\begin{equation}
    \begin{aligned}
        &[P^*, Y_d^*, \beta_{id}^*] = \arg \min_{P, Y_d, \beta_{id}} \quad \beta_{id}\\
        &\text{s. t.} 
        \begin{bmatrix}
            P & P A_{s} - Y_d C_{1s} & PB_s - Y_d D_{1s} & 0\\
            * & P & 0 & C_{2s}^T\\
            * & * & \beta_{id} I & D_{2s}^T\\
            * & * & * & \beta_{id} I
        \end{bmatrix} > 0\\
        &\quad\,\,\,\,\, P > 0
    \end{aligned}
\end{equation}
where the optimal gain is given by $K_s = (P^*)^{-1}Y_d^*$. 

Observe also that the error dynamics for the detectable subspace in (\ref{eq:eid_step2}) can be rewritten as
\begin{equation}
    \begin{multlined}
        e_{id}(k + 1) = (A_{id} - K_{id}C_{id})e_{id}(k) + \\
        \left(
        \begin{bmatrix}
            0 & -D_i^TE_i
        \end{bmatrix}
        - K_{id}
        \begin{bmatrix}
            -I & 0
        \end{bmatrix}
        \right) \tilde v_i(k)
    \end{multlined}
    \label{eq:extracted_ddynamic}
\end{equation}
where $e_{id}(k) \coloneqq D_i^Te(k)$, $\tilde v_i(k) \coloneqq [v_i(k)^T, v_i(k + 1)^T]^T$. 
Then, by selecting $A_s = A_{id}$, $B_s = [0\,\,-D_i^TE_i]$, $C_{1s} = C_{id}$, $C_{2s} = I$, and $D_{1s} = [-I\,\, 0]$ and $D_{2s} = 0$, the systems (\ref{eq:generic_model}) and (\ref{eq:extracted_ddynamic}) become equivalent, leading to optimization problem (\ref{eq:problem_kinf}), thereby concluding the proof.
\end{proof}
\subsection{Design of the stabilizing gain for the undetectable subspace}
The evolution of the derived error in the detectable (\ref{eq:eid_step2}) and undetectable (\ref{eq:error_usubspace}) subspaces indicates that local measurement primarily steers the observer node within the detectable subspace. To operate in the undetectable subspace, a consensus protocol is introduced, governed by the diffusive coupling gains $g_i$. This approach utilizes neighboring data to compensate for the limited knowledge of the internal system state. Specifically, the consensus strategy ensures that each node updates its estimate based on neighboring information until consensus is reached in the undetectable subspace. 
\begin{proposition}
    Given system (\ref{eq:d_system}) under Assumptions \ref{ass:rank_B}, \ref{ass:rank_CB} and \ref{ass:bounded_meas_err} if there exist scalars $g_i^*$ for $i = 1,\dots, m$ solutions of the following semi-definite programming (SDP) optimization problem:
    %the error dynamic in the undetectable subspace (\ref{eq:error_aggr_und}) is bounded by selecting $g_i$ for $i = 1,\dots, m$ as solution of the following optimization problem
    \begin{align}
            &[g^*, \beta_u^*] = \arg \min_{g, \beta_u} \quad \beta_u \label{eq:obj_fun}\\
            &\text{s. t.} \quad M_g \leq \beta_uI\label{eq:sdp_constraint}\\
            & \quad\,\,\,\,\quad\beta_u \leq 0 \label{eq:beta_constraint}
    \end{align}
where 
\begin{gather}
    M_g \hspace{-0.5mm} = \hspace{-0.5mm}
    \begin{bmatrix}
   \left[A_{u}^TA_{u} - \left(A_u^T\Theta(g)\right) - \left(A_u^T \Theta(g) \right)^T - I\right]\vspace{-1mm} & \Theta(g)^T\\
    * & -I
    \end{bmatrix}\\
    \Theta(g) = \left(U^T\left(\operatorname{diag}(g)\mathcal{L}_{g} \otimes I_{n_x}\right)U\right)
\end{gather}
with $g = [g_1,\dots, g_m]^T$, then the error dynamic in the undetectable subspace (\ref{eq:aggr_err_u}) is bounded $\forall k \geq 0$ with error decay rate $\beta_u$.
\end{proposition}

\begin{proof}
    Due to the lower-triangular structure in (\ref{eq:aggr_error}) and the boundedness condition in Assumption \ref{ass:bounded_meas_err}, it suffices to establish the stability of the following subsystem (captured in (\ref{eq:aggr_err_u}))
    \begin{equation}
    U^Te(k+1) = \left[A_{u} - g\, U^T(\mathcal{L}_g \otimes I)U\right] U^Te(k)
    \end{equation}
    which in turn guarantees the boundedness of the error dynamics in the undetectable subspace. We consider the candidate Lyapunov function
\begin{equation}
    V_k =\sum_{i = 1}^m (U^T_ie_i(k))^T(U_i^Te_i(k))
    \label{eq:lypunov_candidate_k}
\end{equation}
which is a positive definite function of the estimation errors. For simplicity of exposition, we have used $V_k \coloneqq V(k)$.  Then, at time $k+1$, we have
\begin{align}
    V_{k+1} \hspace{-1mm}=& \sum_{i = 1}^m U^T_ie_i(k+1))^T(U_i^Te_i(k+1))\nonumber\\
    =& \sum_{i = 1}^m 
    \left[A_{iu}U_i^Te_i(k) - g_iU_i^T\sum_{j=1}^ml_{ij}U_jU_j^Te_j(k)\right]^T\nonumber\\
    &\left[A_{iu}U_i^Te_i(k) - g_iU_i^T\sum_{j=1}^ml_{ij}U_jU_j^Te_j(k)\right]\nonumber\\
    =& \sum_{i = 1}^m 
    (U^T_ie_i(k))^TA_{iu}^TA_{iu}(U^T_ie_i(k))\nonumber\\
    &-\sum_{i = 1}^m \left[\hspace{-1mm}\left(\hspace{-0.5mm}g_i U_i^T\sum_{j=1}^ml_{ij}U_j U_j^Te_j(k)\hspace{-1mm}\right)^T\hspace{-3mm}A_{iu}(U_i^Te_i(k))\right.\nonumber\\
    &\left.-(U^T_ie_i(k))^TA_{iu}^T\left(g_i U_i^T\sum_{j=1}^ml_{ij}U_j U_j^T e_j(k)\right)\right]\nonumber\\
    &+\sum_{i = 1}^m \left[ \left(-g_i U_i^T\sum_{j=1}^ml_{ij}U_j U_j^Te_j(k)\right)^T\right.\nonumber\\
    &\left.\left(-g_i U_i^T\sum_{j=1}^ml_{ij}U_j U_j^Te_j(k)\right)\right]
\end{align}
where the second equality was obtain leveraging the identity derived in  (\ref{eq:error_usubspace}).
By introducing notation $\Theta(g) \coloneqq \left(U^T\left(\operatorname{diag}(g)\mathcal{L}_{g} \otimes I_{n_x}\right)U\right) $, the expression for $V(k+1)$ can be compactly rewritten as
\begin{align}
    V_{k+1}
    % \sum_{i = 1}^m  
    % (U^T_ie_i(k))^TA_{iu}^TA_{iu}(U^T_ie_i(k))\nonumber\\
    % %
    % &- \left(U^T e(k)\right)^T\left(\left(A_u^T \Theta \right) + \left(A_u^T \Theta \right)^T\right)(U^T e(k))\nonumber\\
    % &+\sum_{i = 1}^m \left[ \left(U^T e(k)\right)^T\left(-g_i U_i^T\left(L_{gi} \otimes I_{n_x}\right)\right)^T\left(-g_i U_i^T\left(L_{gi} \otimes I_{n_x}\right)\right) \left(U^T e(k)\right)\right]\\
    %%%
    =&
     (U^Te(k))^TA_{u}^TA_{u}(U^Te(k))\nonumber\\
    &- \left(U^T e(k)\right)^T\left(A_u^T \Theta(g) +  \Theta(g)^TA_u\right)(U^T e(k))\nonumber\\
    &+\left(U^T e(k)\right)^T\Theta(g)^T\Theta(g) \left(U^T e(k)\right)
    \label{eq:lyapunov_candidate_kp1}
\end{align}
For $V_k$ to be a valid Lyapunov function, we require that
\begin{equation}
    V_{k+1} \leq V_k
\end{equation}
along then system trajectories that leads to the following inequality to be satisfied
\begin{equation}
    V_{k+1} - V_k\leq \beta_u
    \label{eq:lyapunov_cond}
\end{equation}
with $\beta_u \in \mathbb{R}$ being a non-positive scalar. Substituting \eqref{eq:lypunov_candidate_k} and \eqref{eq:lyapunov_candidate_kp1} into \eqref{eq:lyapunov_cond} leads to
\begin{align}
    &(U^Te(k))^T \left[A_{u}^TA_{u}- \left(\left(A_u^T \Theta(g) \right) + \left(A_u^T \Theta(g) \right)^T\right)\right.\nonumber\\
    &\left.+\Theta(g)^T\Theta(g)\right] \left(U^T e(k)\right)
    \nonumber\\
    &- \left(U^T e(k)\right)^T\left(U^T e(k)\right)\leq \beta_u
    \label{eq:negative_ieq}
\end{align}
That is equivalent to the following matrix inequality 
\begin{multline}
    \left[\underbrace{A_{u}^TA_{u}- \left(\left(A_u^T \Theta(g) \right) + \left(A_u^T \Theta(g) \right)^T\right) - I}_{M_{11}}\right.\nonumber\\
    \left.-\underbrace{\Theta(g)^T}_{M_{12}} \underbrace{-I^{-1}}_{M_{22}^{-1}}\underbrace{\Theta(g)}_{M_{21}}\right]\leq \beta_u I
    \label{eq:LMI}
\end{multline}
and, by applying Shur's complements, we derive
\begin{equation}
    \begin{bmatrix}
        M_{11} & M_{12}\\
        M_{21} & M_{22}
    \end{bmatrix} \leq \beta_u I
\end{equation}
that is equivalent to the LMI condition (\ref{eq:sdp_constraint}). Minimizing $\beta_u$ in this formulation (subject to non-positivity constraint) yields the optimization problem in (\ref{eq:obj_fun})-(\ref{eq:beta_constraint}) concluding the proof.
\end{proof}
 It is important to note that by minimizing the parameter $\beta_u$, we select, among all feasible diffusive gains $G_i$, those that maximize the decay rate of the Lyapunov function $V(k)$ and thus enhance the convergence properties of the consensus protocol.
 \begin{remark}
     Problem (\ref{eq:obj_fun})-(\ref{eq:beta_constraint}) requires no specific constraints on the graph topology $\mathcal{G}$; in particular, the communication graph does not need to be connected. Numerical tests suggest that the feasibility seems to be solely related to the Mahler measure \citep{C:13,M:60} of $A$ (that is the absolute product of unstable eigenvalues of $A$) and the detectability of the pair $(A, C)$.

 \end{remark}
\section{Numerical Example}
\label{sec:numerical_example}

The node-wise decomposition relies on the triangular structure derived in equations (\ref{eq:decomposition1})-(\ref{eq:decomposition2}). Specifically, this triangular structure arises from the orthogonality condition between subspaces $\mathcal{D}_i$ and $\mathcal{U}_i$, namely $D_i^T U_i = 0$ for $i = 1, \dots, m$. In fact, $\mathcal{D}_i$ is directly determined as the the orthogonal complement of $\mathcal{U}_i$. Moreover, $\mathcal{U}_i$ is an invariant subspace for the matrix $P_i A$ and is contained within $\ker(C_i)$. Consequently, any decomposition that decompose the space $\mathbb{R}^{n_x}$ into two orthogonal subspaces in which one of the two is a $P_i A$-invariant subspace contained within $\ker(C_i)$ can be utilized.

For instance, an alternative viable decomposition is the observability decomposition, where the state space is decomposed into observable and unobservable components. Indeed, it is well-known that for a given matrix pair $(P_i A, C_i)$, the largest $P_i A$-invariant subspace contained within $\ker(C_i)$ is the unobservable subspace associated with the pair $(P_i A, C_i)$ (note that the undetectable subspace is always a subset of the unobservable subspace). Thus, the matrix $D_i$ can be computed as follows:
\begin{equation}
    \operatorname{Im} D_i =\operatorname{Im}\left(
    \begin{bmatrix}
        C_i\\
        C_i P_i A\\
        \vdots\\
        C_i \left(P_iA\right)^{n_x - 1}\\
    \end{bmatrix}^T\right)
\end{equation}
where the matrix on the right hand part of the equality is the observability matrix at node $i$. The matrix $U_i$ is than determined in such a way the matrix $[D_i \quad U_i]$ is an orthonormal matrix.
In this section, we examine two examples based on either the detectability or the observability decompositions in order to assess the effectiveness of the proposed distributed D-UIO observer.

\subsection{Example based on the detectability decomposition}
\begin{figure}[ht!]
	\centering
    \includegraphics[width=\linewidth]{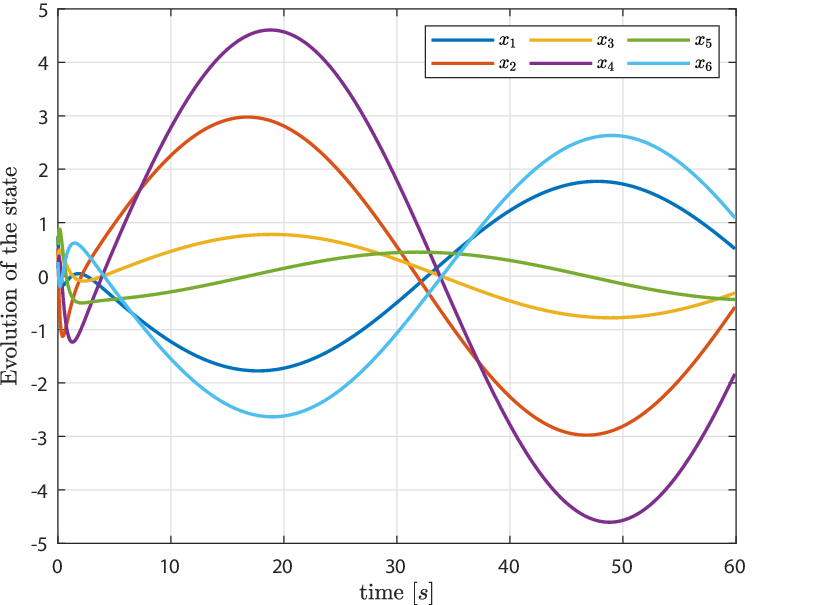}
	\caption{Evolution of the system's state  affected by the disturbance $w(k)$.}
	\label{fig:state_evolution}
\end{figure}

 In the following, the effectiveness of the proposed strategy is analyzed when a node-wise detectability decomposition is adopted. In particular, the LTI system presented in \citep{CW:23} is used as testbed. The model has been discretized using a Zero-Order Hold technique with a sampling time of $t_c = 0.1\,\mathrm{s}$. The system matrices are given below
\begin{gather}
\begin{aligned}
&A = 10^2 \times\\
&\begin{bmatrix}
10.39  & 0      & 0      & 0      & 0      & 0      \\
0.31   & 10.36  & 0      & 0      & 0      & 0      \\
1.59   & 0.67   & 11.06  & 0      & 0      & 0      \\
0.25   & 0      & 0      & 10.58  & 0      & 0      \\
0.01   & 0      & 0      & 0.48   & 11.26  & -0.02  \\
0.00   & 0      & 0      & 0.05   & 2.46   & 11.05
\end{bmatrix}
\end{aligned}\nonumber\\
B  = 
\begin{bmatrix}
0      & 0      & 0.1019 \\
0.1018 & 0      & 0.0015 \\
0.0033 & 0      & 0.0078 \\
0      & 0.1029 & 0.0012 \\
0      & 0.1085 & -0.0001 \\
0      & 0.0120 & 0.1052
\end{bmatrix}, \,
B_w =
\begin{bmatrix}
0\\
0\\
0\\
0\\
0\\
0.1
\end{bmatrix}
\end{gather}
Consider a scenario where a network of five observers, connected in a ring topology is monitoring the evolution of the system. The observers can receive data of the manipulable control input but node $1$ and node $3$ have no access to the third component of the control action. More in detail, by representing the manipulable input matrix as $B=[B^1, B^2, B^3]$, where $B^i$ for $i=1,\dots, n_u$ are the columns of $B$, then the input matrices of the available data input at node $1$ and $3$ are $B_1 = B_3 = [B^1, B^2]$. Each observer is associated with the following sensing capabilities 
\begin{gather}
C_1 = 
\begin{bmatrix}
    1 & 0 & 0 & 0 & 0 & 1 \\
    1 & 1 & 0 & 0 & 0 & 0
\end{bmatrix} 
C_2 = 
\begin{bmatrix}
    1 & 0 & 1 & 0 & 0 & 1 \\
    0 & 1 & 0 & 1 & 0 & 1
\end{bmatrix}\nonumber\\
C_3 = 
\begin{bmatrix}
    1 & 0 & 0 & 0 & 0 & 0 \\
    0 & 0 & 0 & 0 & 0 & 1
\end{bmatrix}
C_4 = 
\begin{bmatrix}
    0 & 0 & 0 & 1 & 0 & 0 \\
    0 & 0 & 0 & 0 & 0 & 1
\end{bmatrix}\nonumber\\
C_5 = 
\begin{bmatrix}
    0 & 0 & 0 & 0 & 1 & 1 \\
    0 & 0 & 0 & 0 & 0 & 1
\end{bmatrix}
\label{eq:sys_matrix}
\end{gather}
 Furthermore, a zero-mean measurement noise $v(k)$ with covariance matrix $Q_v = 1^{-3} I_{n_y}$ and an unknown input $w(k) = 2\sin\left(\frac{2\pi}{T} kt_c\right)$, where $T=60s$ is the simulation time, are considered.
It is important to notice that each sensor is not able to simultaneously isolate the unknown input and reconstruct the entire state alone, indeed the pairs $(P_iA, C_i)$ are not detectable $\forall i\in \mathcal{N}$. To stabilize the system evolution, a control strategy of the form
\begin{equation}
    u(k) = F x(k)
\end{equation}
where $F \in \mathbb{R}^{n_x\times n_u}$ is the feedback control gain computed by solving a linear quadratic regulator problem, is adopted. The evolution of the states is shown in Fig. \ref{fig:state_evolution}.
The control action drives the state evolution towards the origin. However, the unknown input induces persistent oscillations throughout the entire simulation.
\begin{figure}[ht!]
	\centering
		\includegraphics[width=\linewidth]{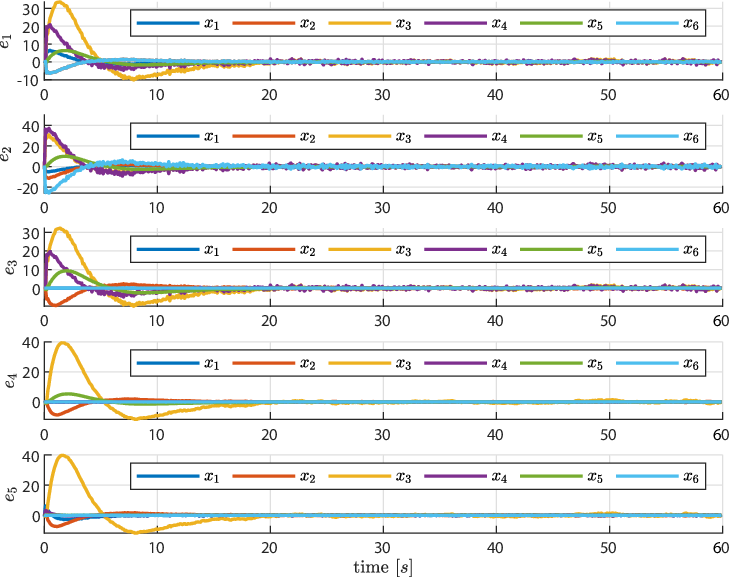}
	\caption{Error evolution for each node of the  $H_{\infty}$ optimal distributed unknown input observer.}
	\label{fig:error_evolutions1}
\end{figure}
Fig. \ref{fig:error_evolutions1} illustrates the error evolution recorded by all the observers. The gains computed by solving the optimization problem (\ref{eq:obj_fun})-(\ref{eq:beta_constraint}) are given by
\begin{equation}
    g = \begin{bmatrix}
     0.4962   & 0.4992  &  0.4976  &  0.4894 &   0.4702
    \end{bmatrix}^T
\end{equation}
Notably, all nodes are able to fully reconstruct the internal state despite the presence of both unknown inputs and measurement errors.
\begin{figure}[ht!]
	\centering
		\includegraphics[width=\linewidth]{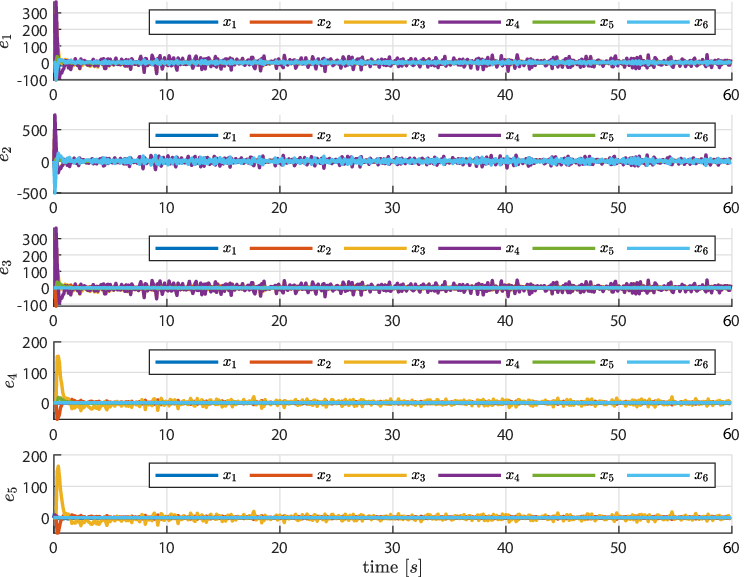}
	\caption{Error evolution for each node recorded when $K_{id}$ are determined with a non-optimal pole-placement strategy.}
	\label{fig:error_evolutions2}
\end{figure}

Figure \ref{fig:error_evolutions2} illustrates the error evolution observed when the stabilizing gains $K_{id}$  for the detectable subspace are determined using a non-optimal pole-placement strategy. While the error remains bounded, the influence of measurement noise is more noticeable.

\subsection{Example based on the observability decomposition}
In the following, the effectiveness of the proposed strategy is analyzed when a node-wise observability decomposition is adopted. Specifically, a simplified LTI system derived from the heat exchange model in multi-zone buildings as presented in \citep{YBRP:22} is considered. The model has been discretized using a Zero-Order Hold technique with a sampling time of $t_c = 60\,\mathrm{s}$. The systems models a building floor divided into nine zones (rooms) by walls with varying heat exchange rates. Heating, ventilation, and air conditioning (HVAC) is installed in three zones (Rooms $2$, $5$, and $6$), while one room (Room $9$) is subject to an unpredictable temperature disturbance $w(k)$. Four observers are deployed at different locations on the floor, with the objective that each observer estimates the temperature of all rooms. 

Furthermore, it is assumed that at most only one HVAC input signal is accessible by each node or observer, while the fourth node or observer is not able to access any of the inputs. Specifically, by representing with $B^1, B^2, \dots, B^4$ the columns of the known input matrix $B$ (here $n_u = m$), the local known input at node $i$ is $u_i(k) = B^iu(k)$. Notice that $B^4$ is equal to the zero column vector.
The state transition matrix, the manipulable input 
%(with entries rounded to one decimal term for space limitations), 
unknown input matrices and output matrices are reported in Appendix A.

\begin{figure}[t]
    \centering
    \begin{subfigure}[]{0.45\linewidth}
        \centering
        \includegraphics[width=1\linewidth]{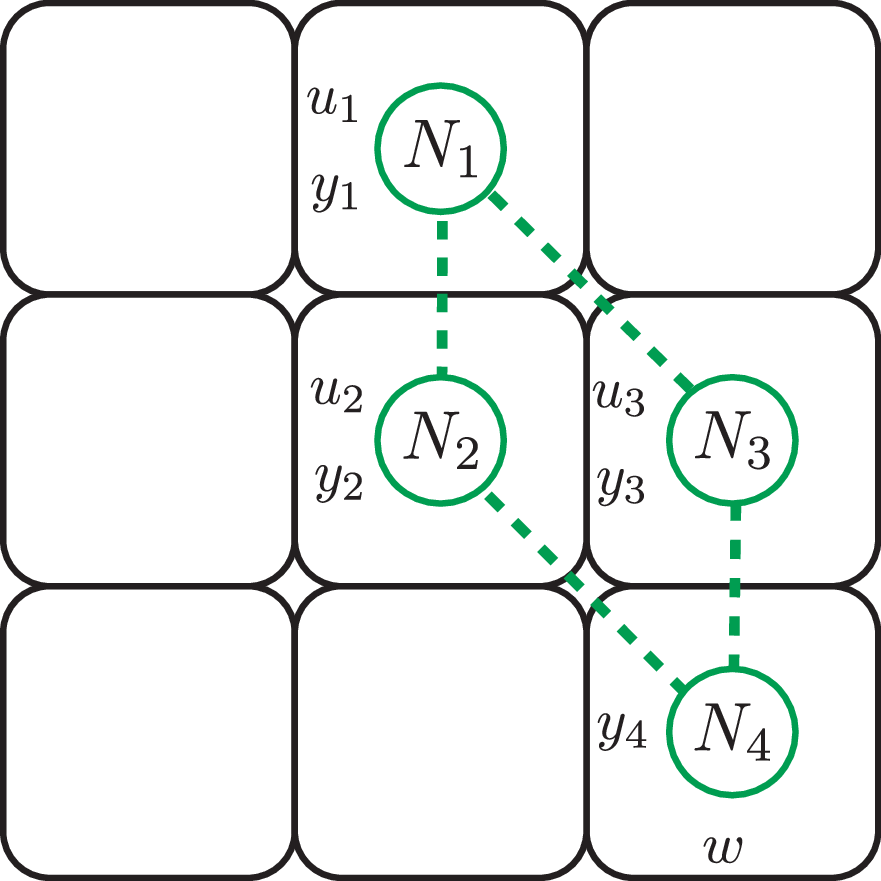}
        \caption{Scenario 1.}
    \end{subfigure}
    ~
    \begin{subfigure}[]{0.45\linewidth}
        \centering
        \includegraphics[width=1\linewidth]{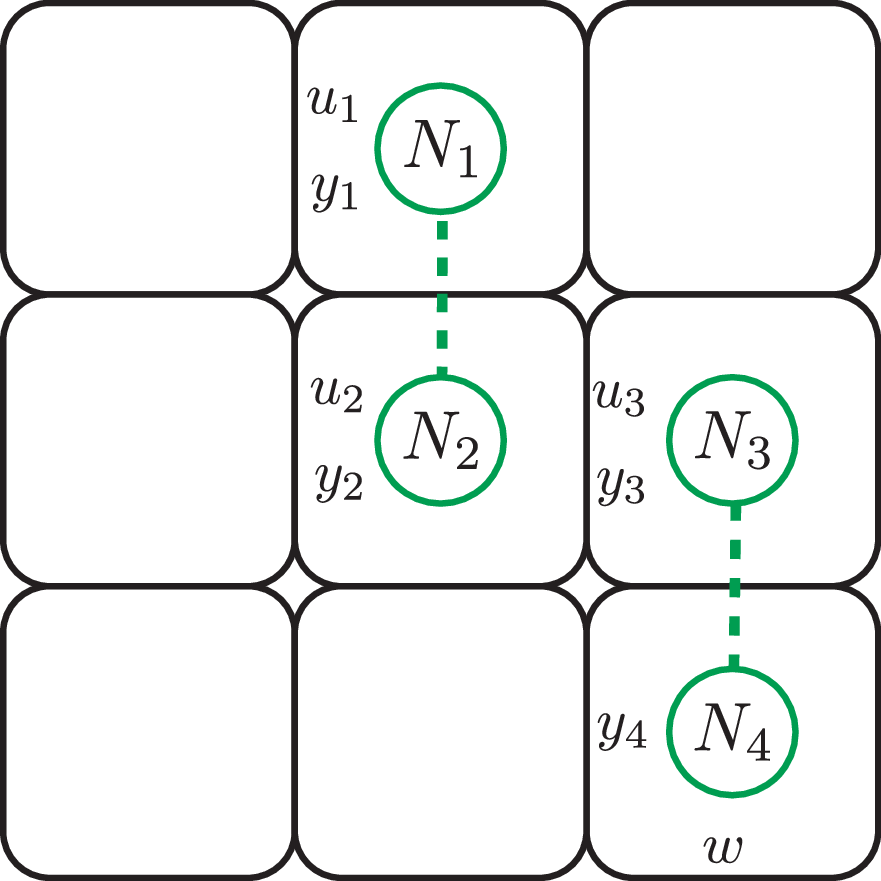}
        \caption{Scenario 2.}
    \end{subfigure}
    \caption{Observers configuration considered in the two scenarios. (a) a connected topology (b) a partitioned topology.}
    \label{fig:scenario_setups_s2}
\end{figure}

It is important to notice that each sensor is not able to simultaneously isolate the unknown input and reconstruct the entire state alone, indeed $\operatorname{rank}(O_i) <n_x$ for $i=1, \dots, m$. We consider two distinct scenarios that differ primarily in the topology of the communication network. In the first scenario, nodes are arranged in a ring topology while the network is not connected and can be partitioned into two sub-graphs in the second. A graphical representation of these setups is provided in Fig. \ref{fig:scenario_setups_s2}.

In both scenarios, a zero-mean measurement noise $v(k)$ with covariance matrix $Q_v = 0.2 I_{n_y}$ and an unknown input $w(k) = 2\sin\left(\frac{2\pi}{T} kt_c\right)$, where $T=6000s$ is the simulation time, are considered.

\begin{figure}[hbt]
	\centering
		\includegraphics[width=\linewidth]{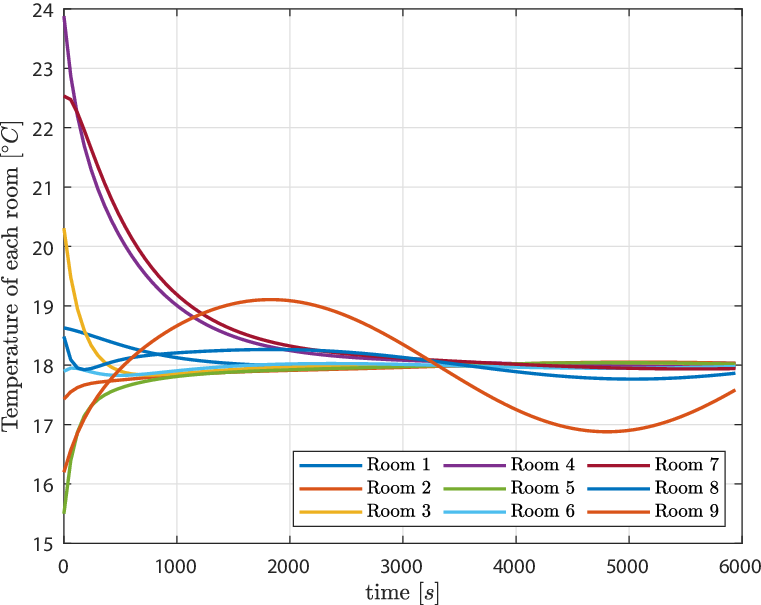}
	\caption{Evolution of the temperature in each of the nine rooms. Room nine is affected by the disturbance $w(k)$.}
	\label{fig:state_evolution_s2}
\end{figure}

A control strategy of the form
\begin{equation}
    u(k) = F\left(x(k) - x_{\text{ref}}\right)
\end{equation}
is considered to steer and maintain the temperature to a reference value, where $F \in \mathbb{R}^{n_u \times n_x}$ is the feedback control gain computed by solving a linear quadratic regulator problem, and $x_{\text{ref}}$ is the reference temperature, set to $18^\circ$. The evolution of the temperature in all nine rooms is shown in Fig. \ref{fig:state_evolution_s2}.
The HVAC system steers the temperature in each room towards the reference temperature of $18^\circ$. However, the unknown input causes the temperature in room $9$ to oscillate between $\sim19^\circ$ and $\sim17^\circ$, indirectly affecting the temperatures in all the other rooms.
\begin{figure}[ht!]
	\centering
		\includegraphics[width=\linewidth]{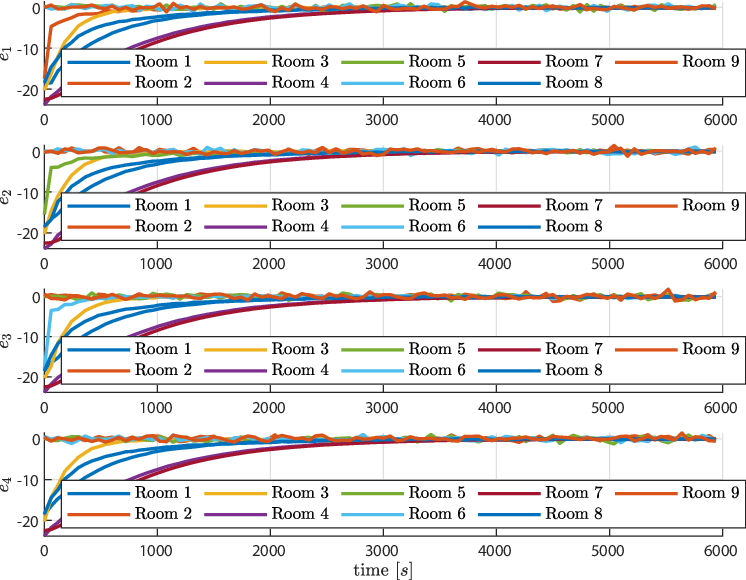}
	\caption{First scenario. Error evolution for each node of the distributed unknown input observer.}
	\label{fig:error_evolutions1_s2}
\end{figure}
Fig. \ref{fig:error_evolutions1_s2} illustrates the error evolution recorded by all the observers in the first scenario. For the connected topology, the gains computed by solving the optimization problem (\ref{eq:obj_fun})-(\ref{eq:beta_constraint}) are given by
\begin{equation}
    g = \begin{bmatrix}
    0.3335   &  0.3348  &  0.3315  &  0.3330
        %0.1387 & 0.1368 & 0.1181 & 0.1205
    \end{bmatrix}^T
\end{equation}
Notably, all nodes are able to fully reconstruct the internal state despite the presence of both an unknown input and measurement error.

\begin{figure}[ht!]
	\centering
		\includegraphics[width=\linewidth]{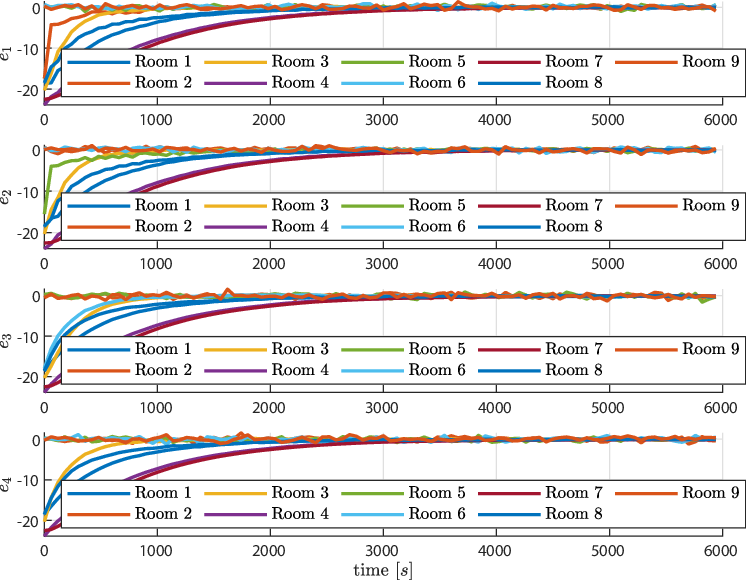}
	\caption{Second scenario. Error evolution for each node of the distributed unknown input observer.}
	\label{fig:error_evolutions2_s2}
\end{figure}
In the second scenario, the communication network is not connected. 
It is interesting to note that each sub-network enjoys the joint observability property; indeed, both pairs $(A, [C_1^T\,\,C_2^T]^T)$ and $(A, [C_3^T\,\,C_4^T]^T)$ are observable. 
The diffusive gains computed by solving problem (\ref{eq:obj_fun})-(\ref{eq:beta_constraint}) are
\begin{equation}
    g = \begin{bmatrix}
    0.6690  &  0.6685 &   0.1434 &  0.1428
    \end{bmatrix}^T
\end{equation}
In this case as well, all the observers are able to fully reconstruct the internal state relying solely on local and neighboring data.

\section{Conclusions}
\label{sec:conclusions}
This work proposed a novel strategy for designing a D-UIO for discrete LTI systems based on node-wise detectability decomposition. The design process was divided into two separate problems: the design of local output injection gains to stabilize the estimation error in the detectable subspace and mitigate the effect of the local measurement error and the design of diffusive gains to stabilize the error in the undetectable subspace. Both problems were addressed through an SDP optimization approach.
The proposed strategy did not require any assumptions on the connectivity of the communication topology. Future research may include the application of the UIO to more realistic scenarios.

% \section*{Acknowledgment}
% This work was partially supported by the European Union - Next Generation EU under the Italian National Recovery and Resilience Plan (NRRP), Mission 4, Component 2, Investment 1.3, CUP E63C22002070006, partnership on “Telecommunications of the Future” (PE00000001 - program “RESTART”).

\bibliographystyle{elsarticle-harv}        % Include this if you use bibtex 
\bibliography{manus}           % and a bib file to produce the 
                                 % bibliography (preferred). The
                                 % correct style is generated by
                                 % Elsevier at the time of printing.
                                 
\section{Appendix A}
This section provides the matrices of the discretized heat exchange model used in the observability decomposition example. In particular, the overall known manipulable input matrix and the known input matrix of the unknown input are
\begin{gather}
B =
% B = \begin{bmatrix}
%         B_1\\
%         B_2\\
%         B_3\\
%         B_4
%     \end{bmatrix} = 
\begin{bmatrix}
0.4 & 5.0 & 0.4 & 0.0 & 0.2 & 0.0 & 0.0 & 0.0 & 0.0 \\
0.0 & 0.2 & 0.0 & 0.2 & 4.9 & 0.1 & 0.0 & 0.5 & 0.0 \\
0.0 & 0.0 & 0.6 & 0.0 & 0.1 & 5.0 & 0.0 & 0.0 & 0.3 \\
0.0 & 0.0 & 0.0 & 0.0 & 0.0 & 0.0 & 0.0 & 0.0 & 0.0
\end{bmatrix}^T\\
    B_w =
\begin{bmatrix}
\mathbf{0}_{8\times 1}\\
0.1
\end{bmatrix}\,
\end{gather}
The state transition matrix is
{\vspace{-5mm}\small
\begin{equation}
\begin{aligned}
&A = 10^{-3}\times\\
&\begin{bmatrix}
844.4 & 114.0 & 8.3   & 23.4  & 5.9   & 0.8   & 2.8   & 0.5   & 0.0   \\
114.0 & 695.7 & 103.6 & 4.4   & 60.5  & 14.0  & 0.5   & 6.4   & 0.8   \\
8.3   & 103.6 & 704.2 & 0.4   & 9.4   & 162.8 & 0.0   & 1.0   & 10.4  \\
23.4  & 4.4   & 0.4   & 721.4 & 61.7  & 1.9   & 174.7 & 11.7  & 0.4   \\
5.9   & 60.5  & 9.4   & 61.7  & 658.8 & 40.4  & 11.9  & 142.1 & 9.3   \\
0.8   & 14.0  & 162.8 & 1.9   & 40.4  & 682.6 & 0.3   & 8.2   & 89.1  \\
2.8   & 0.5   & 0.0   & 174.7 & 11.9  & 0.3   & 763.2 & 44.5  & 2.1   \\
0.5   & 6.4   & 1.0   & 11.7  & 142.1 & 8.2   & 44.5  & 717.0 & 68.7  \\
0.0   & 0.8   & 10.4  & 0.4   & 9.3   & 89.1  & 2.1   & 68.7  & 819.2
\end{bmatrix}
\end{aligned}
\end{equation}
}
The output matrices associated to the $4$ nodes are
\begin{equation}
\begin{aligned}
C_1 &=
\begin{bmatrix}
0 & 0 & 0 & 0 & 1 & 0 & 0 & 0 & 0 \\
0 & 0 & 0 & 0 & 0 & 1 & 0 & 0 & 0 \\
0 & 0 & 0 & 0 & 0 & 0 & 0 & 0 & 1
\end{bmatrix} 
C_2 =
\begin{bmatrix}
0 & 1 & 0 & 0 & 0 & 0 & 0 & 0 & 0 \\
0 & 0 & 0 & 0 & 0 & 1 & 0 & 0 & 0 \\
0 & 0 & 0 & 0 & 0 & 0 & 0 & 0 & 1
\end{bmatrix} \\
C_3 &=
\begin{bmatrix}
0 & 1 & 0 & 0 & 0 & 0 & 0 & 0 & 0 \\
0 & 0 & 0 & 0 & 1 & 0 & 0 & 0 & 0 \\
0 & 0 & 0 & 0 & 0 & 0 & 0 & 0 & 1
\end{bmatrix}
C_4 =
\begin{bmatrix}
0 & 1 & 0 & 0 & 0 & 0 & 0 & 0 & 0 \\
0 & 0 & 0 & 0 & 1 & 0 & 0 & 0 & 0 \\
0 & 0 & 0 & 0 & 0 & 1 & 0 & 0 & 0 \\
0 & 0 & 0 & 0 & 0 & 0 & 0 & 0 & 1
\end{bmatrix}
\end{aligned}
\end{equation}

 \begin{ack}
    This work was partially supported by the European Union - Next Generation EU under the Italian National Recovery and Resilience Plan (NRRP), Mission 4, Component 2, Investment 1.3, CUP E63C22002070006, partnership on “Telecommunications of the Future” (PE00000001 - program “RESTART”).
\end{ack}
\end{document}